\documentclass{svjour2}                    

\usepackage{graphicx}
\usepackage[applemac]{inputenc}
\usepackage{amssymb,amsmath}

\begin{document}

\title{The quantum origin of inertia and the radiation reaction of self-interacting electron}

\author{Peter Leifer}

\institute{Peter Leifer \at
              Haatid, College for Science and Technology, 18123 Afula, Israel  \\
              Tel.: +972-4-6405580\\
              Fax: +972-4-6405576\\
              \email{leifer@bezeqint.net }           
           \and
}

\date{Received: date / Accepted: date}
\maketitle

\begin{abstract}
The internal structure of self-interacting quantum particle like electron is independent on space-time position. Then at least infinitesimal kinematic space-time shift, rotation or boost lead to the equivalent internal quantum state. This assumption may be treated as internal (quantum) formulation of the inertia principle. Dynamical transformation of quantum setup generally leads to deformation of internal quantum state and measure of this deformation may be used as quantum counterpart of force instead of a macroscopic acceleration. The reason of inertia arises, thereby, as a consequence an internal motion of quantum state and its reaction on dynamical quantum setup deformation.

The quantum origin of the inertia has been discussed in this article in the framework of ``eigen-dynamics" of self-interacting relativistic extended quantum electron. Namely, a back reaction of spin and charge ``fast" degrees of freedom on ``slow" driving environment during ``virtual measurement" leads to the appearance of state dependent non-Abelian gauge fields in dynamical 4D space-time. Analysis of simplified dynamics has been applied to the energy-momentum behavior in the relation with runaway solutions of previous models of an electron.
\end{abstract}


\section{Introduction. An ``internal" formulation of the inertia principle}
The fundamental question of the nature of inertial mass is not
solved up to now. Success of Newton's conception of physical force
influencing on a separated body may be explained by the fact that the \emph{geometric counterpart to the force $\vec{F}$  - acceleration $\vec{a}$ in some inertial frame}
was found with the simplest relation $\vec{a}=\frac{\vec{F}}{m}$ to the mass $m$ of a body. The consistent formulation of mechanical laws has been realized in Galilean  inertial systems. The class of the inertial systems contains (by a convention) the one unique inertial system - the system of remote stars and any reference frame moving with constant velocity relative these remote stars. Then, on the abstract mathematical level arose a ``space" - the linear Euclidean space with appropriate vector operations on forces, momenta, velocities, etc. General relativity and new
astronomical observations concerning accelerated expansion of Universe show that all these constructions are only a good approximation, at best.

The line of Galileo-Newton-Mach and Einstein (with serious reservations about  conception of the ``space") argumentations made accent on some absolute global reference frame associated with the system of remote stars. This point of view looks as absolutely necessary for the classical formulation of the inertial principle itself. This fundamental principle has been formulated, say, ``externally", i.e. as if one looks on some massive body perfectly isolated from rest Universe. In such approach only ``mechanical" state of relative motion of bodies expressed by their coordinates in space has been taken into account. Nevertheless, Newton clearly saw some weakness of such approach. His famous example of rotating bucket with water shows that there is an absolute motion since the water takes on a concave shape in any reference frame. Here we are very close to different - ``internal" formulation of the inertia principle and, probably, to understanding  the quantum nature of inertial mass. Namely, the ``absolute motion" of a body should be turned towards not outward, to distant stars, but inward -- to the deformation of the body. This means that external force not only changes the inertial character of its motion:  body accelerates, moreover -- the body deforms.

Two aspects of a force action: acceleration relative inertial reference frame and
deformation of the body are very important already on the classical level as it has been shown by Newton's bucket rotation. The second aspect is especially important for quantum ``particles" since the acceleration requires the point-like localization in space-time; such localization is, however, very problematic in quantum theory.
Nevertheless, almost all discussions in foundations of quantum theory presume that
space-time structure is close with an acceptable accuracy to the Minkowski
geometry and may be used without changes in quantum theory up to Planck's
scale or up to topologically different space-time geometry of string theories.
Under such approach, one loses the fact that space-time relationships and
geometry for quantum objects should be reformulated totally \emph{at any space-time distance} since from the quantum point of view such fundamental dynamical variables as ``time-of-arrival" of Aharonov \textit{et al.} (1998) and a position operator of Foldy \& Wouthuysen (F.-W.) (1950) and Newton \& Wigner (1949) representations are state-dependent (Leifer, 1997,1998,2004,2005,2007,2009,2010). Therefore space-time itself should be built in the frameworks of a new ``quantum geometry".

In such a situation, one should make accent on the second aspect of the force action -- the body deformation. In fact, microscopically, it is already  a \emph{different body} with different temperature, etc., since the state of body is changed (Leifer, 1988). In the case of inertial motion one has the opposite situation -- the internal state of the body does not change, i.e. body is self-identical during inertial space-time motion. In fact this is the basis of all classical physics. Generally, space-time localization being treated as ability of coordinate description of an object in classical relativity closely connected with operational
identification  of ``events" (see (Einstein, 1905)). It is tacitly assumed that all classical objects (frequently represented by material points) are
self-identical and they cannot disappear during inertial motion because of the energy-momentum conservation law. The inertia law of Galileo-Newton ascertains this self-conservation ``externally". But objectively this means that \emph{physical state of body (temporary in somewhat indefinite sense) does not depend on the choice of the inertial reference frame}. One may accept this statement as an ``internal" formulation of the inertial law that should be of course formulated mathematically. I put here some plausible reasonings leading to such a formulation.

Up to now the localization problem of quantum systems in the space-time is connected
in fact with the fundamental classical notion of potential energy and force.
Einstein and Schr\"odinger already discussed the inconsistency of usage such
purely classical notions together with the quantum law of motion and the concepts
of ``particle" and ``acceleration" as well (see one of the letter of Einstein
to Schr\"odinger (Einstein, 1950), and the article (Einstein, 1953)). But
these messages are left almost without attention by the physical community.
Now one should pay the bill.

Newton's force is the physical reason for the \emph{absolute} change of the character of a body motion expressed by \emph{acceleration} serves as geometric
counterpart to the force (curvature of the world line in Newtonian space and time
is now non-zero). However there is no adequate geometric notion in quantum
theory since, for example, the notion of trajectory of quantum system in space-time   was systematically banned. The energy of interaction
expressed by a Hamiltonian $H_{int}$ is in some sense an analoge of a classical force. Generally, this interaction leads to the absolute change (deformation) of the quantum state (Leifer, 1998). Notice, quantum state is in fact the state of motion
(Dirac,1928). Such motion takes the place in a state space modeled
frequently by some Hilbert space $\mathcal{H}$. But there is no geometric
counterpart to $H_{int}$ in such functional space. In order to establish the geometric counterpart to $H_{int}$ it is useful
initially to clarify the important question: what is the quantum content
of a classical force, if any?

Let me use a small droplet of mercury as simple example of macroscopic system. The free droplet of mercury is in the state
of the inertial motion ``whether it be of rest, or of moving uniformly forward in a
straight line".  In fact this statement means that physical states of the droplet (its internal degrees of freedom) being in inertial motion are physically non-distinguishable.  The force applied to the droplet breaks the inertial character of its motion and deforms its surface, changes its surface tension temperature, etc. In fact an external force perturbs
Goldstone's modes supporting the droplet as a macro-system (Umezawa, Matsumoto \& Tachiki, 1982) and micro-potentials acting on any internal quantum particle, say, electrons inside of the droplet. It means that quantum states and their deformations may serve as a ``detector" of the ``external force" action on the droplet.

Therefore it is reasonable to use quantum state deformations as a counterpart instead of classical acceleration, since generally acceleration depends on mass, charge, etc., that is impossible to establish a pure space-time geometrically invariant counterpart of a classical force independent on a material body. Only a classical gravitation force may be geometrized assuming the gravitation field may be replaced locally by an accelerated reference frame since in the general relativity (GR) the gravitation field and accelerated reference frame are locally non-distinguishable.
There is, however, a more serious reason why space-time acceleration cannot serves as the completely robust geometric counterpart of the force.

The physical state of the droplet freely falling in the gravitation field
of a star is non-distinguishable from the physical state of the droplet in
an remote from stars area. It means that from the point of view of the ``physical state" of the droplet, the class of the inertial systems may be
supplemented by a reference frame freely falling in a gravitation field.
Therefore macroscopic space-time acceleration cannot serve as a discriminator of physical state of body. Thus, instead of choosing, say, the system of distant stars as an ``outer" absolute reference frame
(Einstein, 1916) the deformation of quantum state of some particle of the droplet may be used. It means that the deformation of quantum motion in quantum state space serves as an ``internal detector" for ``accelerated" space-time motion.
This deformation being discussed from the quantum point of view gives the alternative way for the connection of $H_{int}$ action with a \emph{new geometric counterpart of interaction - coset structure of the quantum state space}.
It means that instead of absolute external reference frame of remote stars
one may use ``internal", in fact a quantum reference frame (Leifer, 2010,2009,1997,2007,2004,1998,2005).

There is different description of accelerated electron in the thermalized vacuum (Unruh, 1976), (Bell \& Leinaas, 1983),  (Bell\& Leinaas, 1986), (Myhrvold, 1985). Avoiding detailed analysis I should note that any variant of
such description uses the interaction of ``detector" with quantum field embedded in single space-time where the acceleration has absolute sense. Indeed, Unruh temperature $T_U = \frac{\hbar a}{2 \pi c k_B}$ is proportional to the acceleration $a$. Relative what reference frame? Say, freely falling electron is definitely ``accelerated" relative remote stars, but its quantum state should be the same as the quantum state of ``free" electron in remote from masses area. Therefore, I have assumed that the equivalence of pure quantum state of single free and freely falling electron (Leifer, 2009). Then the \emph{deformations of quantum motion generated by the coset action in the quantum state space will be used as an invariant counterpart of a ``quantum force" applied to self-interacting electron in dynamical space-time}. It means nothing but in the developing theory \emph{a distance between quantum states in the state space should replace a distance between ``bodies" in space-time as the primary geometric notion}. Thus, the mathematical formulation of the quantum inertia law requires the intrinsic unification of relativity and quantum principles.

\section{Intrinsic unification of relativity and quantum principles}
The mentioned above the localization problem in space-time and connected with this deep difficulties of divergences in quantum field theory (QFT) insist to find a new primordial quantum element instead of the classical ``material point". Such element is the quantum motion - the quantum state of a system (Dirac, 1928). Quantum states of single quantum particles may be represented by vectors $|\Psi>, |\Phi>,...$ of linear functional Hilbert space $\mathcal{H}$ with finite or countable number of dimensions. It is very important to note that the correspondence between quantum state and its vectors representation in $\mathcal{H}$ is not isomorphic. It is rather homomorphic, when a full equivalence class of proportional vectors, so-called rays $\{\Psi\}= z |\Psi>$, where $z \in \mathcal{C} \setminus \{0\}$ corresponds to one quantum state. The rays of quantum states may be represented by points of complex projective Hilbert space $CP^{\infty}$ or its finite dimension subspace $CP(N-1)$. Points of $CP(N-1)$ represent generalized coherent states (GCS) that will be used thereafter as fundamental physical concept instead of ``material point". This space will be treated as the space of \emph{``unlocated quantum states"} in the analog of the \emph{``space of unlocated shapes"} of (Shapere \& Wilczek, 1989).
The problem we will dealing with is the lift the quantum dynamics from $CP(N-1)$ into the \emph{space of located quantum states}. That is, the difference with the Shapere \& Wilczek construction is that not a self-deformation of 3D-shapes should be represented by motions of a spatial reference frame but the unlocated quantum states should be represented by the motions of ``field-shell" in dynamical space-time.

Two simple observations may serve as the basis of the intrinsic unification of relativity and quantum principles. The first observation concerns interference of quantum amplitudes in a fixed quantum setup.

A. The linear interference of quantum amplitudes shows the symmetries
relative space-time transformations of whole setup. This interference has
been studied in ``standard" quantum theory. Such symmetries reflects, say,
the \emph{first order of relativity}: the physics is same if any
\emph{complete setup} subject (kinematical,
not dynamical!) shifts, rotations, boosts as whole in single Minkowski
space-time. According to our notes given above one should add to this list a freely falling quantum setup (super-relativity).

The second observation concerns a dynamical ``deformation" of some quantum setup.

B. If one dynamically changes the setup configuration or its ``environment", then the amplitude of an event will be generally changed. Nevertheless there is a different type of tacitly assumed symmetry that may be formulated on the intuitive
level as the invariance of physical properties of ``quantum particles",
i.e. the invariance of their quantum numbers like mass, spin, charge,
etc., relative variation of quantum amplitudes. This means that properties of, say, physical electrons in two different setups $S_1$ and $S_2$ are the same.

\emph{One may postulate that the invariant content of this physical properties may be kept if one makes the infinitesimal variation of some
``flexible quantum setup"  reached by a small variation of some
fields by adjustment of tuning devices.}

There is an essential technical approach capable to connect super-relativity and the quantum inertia law. Namely, a new concept of local dynamical variable (LDV) (Leifer, 2004) should be introduced for the realization of infinitesimal variation of a ``flexible quantum setup". This construction is naturally connected with methods developed in studying geometric phase (Berry, 1989). I seek, however, conservation laws for LDV's in the quantum state space.

\section{Invariant classification of quantum motions}
The idealized separation of a quantum system from the rest of Universe and its cyclic evolution under parameters variation of a Hamiltonian in adiabatic and beyond
the adiabatic approximation last time mostly discussed in the framework of geometric phase (Berry, 1989). (Aharonov \& Anandan 1987,1988) found
non-adiabatic generalization of Berry construction interesting in following important direction. Aharonov \& Anandan used the general setting for the geometric phase in terms of connection in fibre bundles over projective Hilbert space. Such approach is important since instead of external parameters of the Hamiltonian, the local projective coordinates of a quantum state itself have been used. For us will be interesting just the invariant classification of quantum motions (Leifer, 1997). These invariant classifications are the quantum analog of classical conditions of the inertial and accelerated motions. They are rooted into the global geometry of the dynamical group $SU(N)$ manifold. Namely, the geometry of $G=SU(N)$, the isotropy group $H=U(1)\times U(N-1)$ of the pure quantum state, and the coset $G/H=SU(N)/S[U(1)\times U(N-1)]=CP(N-1)$ geometry, play an essential role in classification of quantum state motions (Leifer, 1997).  I assumed that dynamical group of pure internal quantum states $|\Psi>$ of ``elementary" quantum particle like electron is $G=SU(N)$ then the isotropy group of $|\Psi>$  in co-moving reference frame is $H_{|\Psi>}=\{g \in G: g|\Psi>=|\Psi>\}=S[U(1) \times U(N-1)]$ and the coset space $G/H_{|\Psi>}=SU(N)/S[U(1) \times U(N-1)]$ will be diffeomorphic to the ray space $CP(N-1)$ that may be treated as state space of internal quantum degrees of freedom. Their self-interacting ``eigen-dynamics" will be studied.

What is the physical reason of application $SU(N)$ group geometry to the state of motion of quantum particles? Just because we try study the motion of internal degrees of freedom in the space of \emph{``unlocated quantum states"} $CP(N-1)$.
This is the simplest and, probably, the fundamental case is the case of ``isolated", ``free" but self-interacting quantum particles in pure quantum state whose \emph{inertial motions} could be geometrically analyzed without any reference to the external interaction with a ``second particle". Geometric analysis of quantum state dynamics requires the introduction of a new quantum LDV's and a new method of their ``dynamical identification" by a comparison.

I introduce the LDV's corresponding to the internal $SU(N)$ dynamical group symmetry of quantum states and its breakdown. These LDV's may be naturally expressed in terms of the following local coordinates $\pi^i_{(j)}$
\begin{eqnarray}
\pi^i_{(j)}=\begin{cases}
\frac{\psi^i}{\psi^j},& if \quad 1 \leq i < j \cr
\frac{\psi^{i+1}}{\psi^j} & if \quad  j \leq i < N
\end{cases}
\end{eqnarray}\label{11}
since $SU(N)$ acts effectively only on the space of rays, i.e. on the classes of equivalence of quantum states differentiating by a non-zero complex multiplier. LDV's will be represented by linear combinations of $SU(N)$ generators in local coordinates of $CP(N-1)$ equipped with the Fubini-Study metric
\begin{equation}
G_{ik^*} = [(1+ \sum |\pi^s|^2) \delta_{ik}- \pi^{i^*} \pi^k](1+
\sum |\pi^s|^2)^{-2}.
\end{equation}\label{2}
Hence the internal dynamical
variables and their norms should be state-dependent, i.e. local in
the state space. These local dynamical variables realize
a non-linear representation of the unitary global $SU(N)$ group in
the Hilbert state space $C^N$. Namely, $N^2-1$ generators of $G =
SU(N)$ may be divided in accordance with the Cartan decomposition:
$[B,B] \in H, [B,H] \in B, [B,B] \in H$. The $(N-1)^2$ generators
\begin{eqnarray}
\Phi_h^i \frac{\partial}{\partial \pi^i}+c.c. \in H,\quad 1 \le h
\le (N-1)^2
\end{eqnarray}\label{3}
of the isotropy group $H = U(1)\times U(N-1)$ of the ray and $2(N-1)$ generators
\begin{eqnarray}
\Phi_b^i \frac{\partial}{\partial \pi^i} + c.c. \in B, \quad 1 \le b
\le 2(N-1)
\end{eqnarray}\label{4}
are the coset $G/H = SU(N)/S[U(1) \times U(N-1)]$ generators
realizing the breakdown of the $G = SU(N)$ symmetry of the generalized coherent states (GCS's). Here $\Phi^i_{\sigma}, \quad 1 \le \sigma \le N^2-1 $
are the coefficient functions of the generators of the non-linear
$SU(N)$ realization. They give the infinitesimal shift of the
$i$-component of the coherent state driven by the $\sigma$-component
of the unitary  field $\exp(i\epsilon \lambda_{\sigma})$ rotating by the
generators of $Alg SU(N)$ and they are defined as follows:
\begin{equation}
\Phi_{\sigma}^i = \lim_{\epsilon \to 0} \epsilon^{-1}
\biggl\{\frac{[\exp(i\epsilon \lambda_{\sigma})]_m^i \psi^m}{[\exp(i
\epsilon \lambda_{\sigma})]_m^j \psi^m }-\frac{\psi^i}{\psi^j} \biggr\}=
\lim_{\epsilon \to 0} \epsilon^{-1} \{ \pi^i(\epsilon
\lambda_{\sigma}) -\pi^i \},
\end{equation}\label{5}
(Leifer, 1997, 2009). Then each of the $N^2-1$ generators
may be represented by vector fields comprised by the coefficient
functions $\Phi_{\sigma}^i$ contracted with corresponding partial derivatives
$\frac{\partial }{\partial \pi^i} = \frac{1}{2}
(\frac{\partial }{\partial \Re{\pi^i}} - i \frac{\partial }{\partial
\Im{\pi^i}})$ and $\frac{\partial }{\partial \pi^{*i}} = \frac{1}{2}
(\frac{\partial }{\partial \Re{\pi^i}} + i \frac{\partial }{\partial
\Im{\pi^i}})$.

\section{Affine state-dependent gauge fields }
The anholonomy of the wave function arose due to slowly variable
environment was widely discussed by Berry and many other authors in the
framework of so-called geometric phases (Berry, 1989). It is clear that
now we deal with different problem: \emph{Berry made accent on variation
of wave function during finite cyclic evolution whereas for us interesting the quantum invariants of infinitesimal variation of the quantum setup}.

The geometric phase is an intrinsic property of the family
of eigenstates. There are in fact a set of  local dynamical variables (LDV)
that like the geometric phase
intrinsically depends on eigenstates. For us will be interesting
vector fields
$\xi^k(\pi^1_{(j)},...,\pi^{N-1}_{(j)}): CP(N-1)\rightarrow \mathcal{C}$
associated with the reaction of quantum state $\pi^i_{(j)}$ on the action of internal ``unitary field" $\exp(i\epsilon \lambda_{\sigma})$ given by $\Phi^i_{\sigma}$.

In view of the future discussion of infinitesimal unitary transformations, it is useful to compare \emph{velocity} of variation of the Berry's phase
\begin{equation}
\dot{\gamma}_n(t) = -\textbf{A}_n(\textbf{R})\dot{\textbf{R}},
\end{equation}\label{7}
where $\textbf{A}_n(\textbf{R})=\Im<n(\textbf{R})| \nabla_{\textbf{R}} n(\textbf{R})>$
with the affine parallel transport of the vector field $\xi^k(\pi^1,...,\pi^{N-1})$
given by the equations
\begin{equation}
\frac{d \xi^i}{d\tau}=-\Gamma^i_{kl}\xi^k \frac{d \pi^l}{d\tau}.
\end{equation}\label{8}
The parallel transport of Berry is similar but it is not identical to the affine parallel transport. The last one is the fundamental because this agrees with Fubini-Study ``quantum metric tensor" $G_{ik^*}$ in the base manifold $CP(N-1)$.
The affine gauge field given by the connection
\begin{eqnarray}
\Gamma^i_{mn} = \frac{1}{2}G^{ip^*} (\frac{\partial
G_{mp^*}}{\partial \pi^n} + \frac{\partial G_{p^*n}}{\partial
\pi^m}) = -  \frac{\delta^i_m \pi^{n^*} + \delta^i_n \pi^{m^*}}{1+
\sum |\pi^s|^2}.
\end{eqnarray}\label{3}
is of course more close to the Wilczek-Zee
non-Abelian gauge fields (Wilczek \& Zee, 1984) where the Higgs potential has been replaced by the affine gauge potential (3) whose shape is depicted in Fig. 1.
\begin{figure}[h]
\begin{center}
    \includegraphics[width=4in]{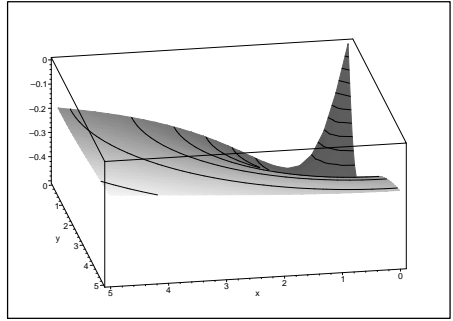}\\
  \caption{The shape of the gauge potential associated with the affine connection in CP(1): $\Gamma=-2\frac{|\pi|}{1+|\pi|^2},
\pi=x+iy. $}\label{fig.1}
  \end{center}
  \end{figure}
It is involved in the affine parallel transport of LDV's (Leifer, 2005,2007) which agrees with the Fubini-Study metric (2).

The transformation law of the connection form
$\Gamma^i_k = \Gamma^i_{kl} d \pi^l$ in $CP(N-1)$ under the differentiable transformations of local coordinates
$\Lambda^i_m=\frac{\partial \pi^i}{\partial \pi^{'m}}$ is as follows:
\begin{eqnarray}
\Gamma'^i_k = \Lambda^i_m\Gamma^m_j \Lambda^{-1j}_k+d\Lambda^i_s \Lambda^{-1s}_k.
\end{eqnarray}\label{10}
It is similar but not identical to well known transformations of non-Abelian fields.
The affine Cartan's moving reference frame takes here the place of ``flexible quantum setup", whose motion refers to itself with infinitesimally close coordinates. Thus we will be rid us of necessity in ``second particle" (Anandan \& Aharonov, 1988), (Leifer, 1997,2007,2009) as an external reference frame. Such construction perfectly fits for the quantum formulation of  the quantum inertia principle (Leifer, 2010) since the affine parallel transport of energy-momentum vector field in $CP(N-1)$ expresses the self-conservation and conditions of stability of, say, electron.

\section{State space and dynamical space-time}
How to lift up the dynamics from the space of \emph{``unlocated quantum states"} $CP(N-1)$ into the DST or quantum state space cum location? The ``inverse representation" of unitary group $SU(N)$ whose action in the space of internal degrees of freedom $CP(N-1)$ should be realized by a ``field shell" motions in dynamical space-time (DST) has been proposed (Leifer, 2010,2009,1997,2007,2004).  Thereby, the space-time degrees of freedom and the space-time geometry itself should be derived in order to describe the lump of energy-momentum distribution in the ``field shell" wrapping internal degrees of freedom.

Let assume that one has, say, a quantum ``free" electron with charge and spin and its quantum state is given by the local projective coordinates $(\pi^1, \pi^2, \pi^3)$.
Dynamical structure of quantum electron should be accepted seriously since now the inertia principle refers just to internal quantum state without evident reference to space-time coordinates. Hence one need initially to deal with the dynamics of quantum degrees of freedom in $CP(3)$. Notice, I am not intended here to establish the connection of this model with the Standard Model. This will be a next step of the investigation.

The distance between two quantum states of electron in $CP(3)$ given by the  Fubibi-Study invariant interval $dS_{F.-S.}=G_{ik*}d\pi^i d\pi^{k*}$. The speed of the interval variation is given by the equation
\begin{eqnarray}
(\frac{dS_{F.-S.}}{d\tau})^2=G_{ik*}\frac{d\pi^i}{d\tau}\frac{d\pi^{k*}}{d\tau}
=\frac{c^2}{\hbar^2}G_{ik*}(\Phi^i_{\mu}P^{\mu})(\Phi^{k*}_{\nu}P^{\nu*})
\end{eqnarray}\label{115}
relative ``quantum proper time" $\tau$ where energy-momentum vector field $P^{\mu}(x)$ obeys field equations that will be derived later. This internal dynamics should be expressed in space-time coordinates $x^{\mu}$ assuming that variation of coordinates $\delta x^{\mu}$ arise due to the transformations of Lorentz reference frame ``centered" about covariant derivative $\frac{\delta P^{\nu}}{\delta \tau}=\frac{\delta x^{\mu}}{\delta \tau}  (\frac{\partial P^{\nu}}{\partial x^{\mu}} + \Gamma^{\nu}_{\mu \lambda}P^{\lambda})$ in dynamical space-time (DST). Such  procedure may be called ``inverse representation" (Leifer, 2010,2009) since this intended to represent quantum motions in $CP(3)$ by ``quantum Lorentz transformation" in DST as will be described below.

Local Lorentz frame will be built on the basis of the qubit spinor $\eta$ whose components may be locally (in $CP(3)$) adjusted by ``quantum boosts" and ``quantum rotations" so that the velocity of the spinor variation coincides with velocity variation of the Jacobi vector fields: tangent Jacobi vector field $\eta^0=J_{tang}(\pi)=(a_i\tau + b_i)U^i(\pi)$ giving initial frequencies traversing the geodesic and the initial phases, and the normal Jacobi vector field $\eta^1=J_{norm}(\pi)=[c_i \sin(\sqrt{\kappa}\tau)+d_i \cos(\sqrt{\kappa}\tau)]U^i(\pi)$ showing deviation from one geodesic to the another one (Besse, 1978). Thereby, two invariantly separated motions have been taken into account:

1)``free motion" of spin/charge degrees of freedom \emph{along energy-momentum tangent vector to geodesic line in CP(3)} (oscillation of massive mode in the vicinity of a minimum of the affine potential across its valley) generated by the coset transformations $G/H=SU(4)/S[U(1) \times U(3)] = CP(3)$ and,

2) deviation of geodesic line motion \emph{in the direction of normal Jacobi vector field transversal to reference geodesic line} (oscillation of massless mode along the valley of the affine potential) generated by the isotropy group $H=U(1) \times U(3)$.

The sufficiency condition of the functional extremum has been applied here to the action for the self-interacting electron. Namely, variation of geodesics in $CP(3)$ represented by the Jacobi fields may be written as the Hamilton equations (Sternberg, 1964) for energy-momentum taking the place of the gauge potentials in DST. This field should serve as a generalization of electromagnetic field since Jacobi equations for variation of geodesics in the symmetric space $CP(N-1)$ with constant sectional curvature $\kappa$ is in fact the equations for the coupled harmonic oscillators. The sufficiency conditions use the concepts of \emph{second Lagrangian, second Hamiltonian and second extremal} (Young, 1969) without, however, evident relations to the method of second quantization. 

The sectional curvature $\kappa$ of $CP(3)$ is not specified yet.  This ``measurement" means that in the DST only deviations from the geodesic motion is ``observable" due to electromagnetic-like field surrounding electron. This field may be mapped onto DST if one assumes that transition from one GCS of the electron to another is accompanied by dynamical transition from one Lorentz frame to another. Thereby, infinitesimal Lorentz transformations define small DST due to coordinate variations $\delta x^{\mu}$.

It is convenient to take Lorentz transformations in the following form
\begin{eqnarray}\label{14}
ct'&=&ct+(\vec{x} \vec{a}_Q) \delta \tau \cr
\vec{x'}&=&\vec{x}+ct\vec{a}_Q \delta \tau
+(\vec{\omega}_Q \times \vec{x}) \delta \tau
\end{eqnarray}
where I put for the parameters of quantum acceleration and rotation the definitions $\vec{a}_Q=(a_1/c,a_2/c,a_3/c), \quad
\vec{\omega}_Q=(\omega_1,\omega_2,\omega_3)$ (Misner,  Thorne, Wheeler, 1973) in order to have
for $\tau$ the physical dimension of time. The expression for the
``4-velocity" $ V^{\mu}$ is as follows
\begin{equation}\label{15}
V^{\mu}_Q=\frac{\delta x^{\mu}}{\delta \tau} = (\vec{x} \vec{a}_Q,
ct\vec{a}_Q  +\vec{\omega}_Q \times \vec{x}) .
\end{equation}
The coordinates $x^\mu$ of imaging point in dynamical space-time serve here merely for the parametrization of the energy-momentum distribution in the ``field
shell'' described by quasi-linear field equations (Leifer, 2009,2010) that will be derived below.

Any two infinitesimally close spinors $\eta$ and $\eta+\delta
\eta$ may be formally connected with infinitesimal $SL(2,C)$ transformations
represented by ``Lorentz spin transformations
matrix'' (Misner,  Thorne, Wheeler, 1973)
\begin{eqnarray}\label{43}
\hat{L}=\left( \begin {array}{cc} 1-\frac{i}{2}\delta \tau ( \omega_3+ia_3 )
&-\frac{i}{2}\delta \tau ( \omega_1+ia_1 -i ( \omega_2+ia_2)) \cr
-\frac{i}{2}\delta \tau
 ( \omega_1+ia_1+i ( \omega_2+ia_2))
 &1-\frac{i}{2}\delta \tau( -\omega_3-ia_3)
\end {array} \right).
\end{eqnarray}
Then ``quantum accelerations" $a_1,a_2,a_3$ and ``quantum angular velocities"
$\omega_1,
\omega_2, \omega_3$ may be found in the linear approximation from
the equation $\delta \eta = \hat{L} \eta-\eta$ and from
the equations for the velocities $\xi$ of $\eta$ spinor variations
expressed in two different forms:
\begin{eqnarray}
\hat{R}\left(
  \begin{array}{cc}
    \eta^0  \cr
    \eta^1
  \end{array}
\right) =
\frac{1}{\delta \tau}(\hat{L}-\hat{1})\left(
  \begin{array}{cc}
    \eta^0  \cr
    \eta^1
  \end{array}
\right) = \left(
  \begin{array}{cc}
    \xi^0 \cr
    \xi^1
  \end{array}
\right)
\end{eqnarray}\label{44}
and
\begin{eqnarray}
 \left(
  \begin{array}{cc}
    \xi^0 \cr
    \xi^1
  \end{array}
\right) =
\left(
  \begin{array}{cc}
    \frac{\delta \{(a_i\tau + b_i)U^i(\pi)\}}{\delta \tau}  \cr
    \frac{\delta \{[c_i \sin(\sqrt{\kappa}\tau)+d_i \cos(\sqrt{\kappa}\tau)]U^i(\pi)\}}{\delta \tau}
  \end{array}
\right)
\end{eqnarray}
Since $CP(3)$ is totally geodesic manifold  (Kobayashi \& Nomizu, 1969),
each geodesic belongs to some $CP(1)$ parameterized by the single complex variable
$\pi=e^{-i\phi} \tan(\theta/2)$. Then the
tangent vector field $U(\pi) = \frac{\delta \pi}{\delta \tau}=
\frac{\partial \pi}{\partial \theta}\frac{\delta \theta}{\delta \tau}+
\frac{\partial \pi}{\partial \phi}\frac{\delta \phi}{\delta \tau}$, where
\begin{eqnarray}\label{45}
\frac{\delta \theta}{\delta \tau}=-\omega_3\sin(\theta)-((a_2+\omega_1)\cos(\phi)+
(a_1-\omega_2)\sin(\phi))\sin(\theta/2)^2 \cr
-((a_2-\omega_1)\cos(\phi)+
(a_1+\omega_2)\sin(\phi))\cos(\theta/2)^2; \cr
\frac{\delta \phi}{\delta \tau}=a_3+(1/2)(((a_1-\omega_2)\cos(\phi)-
(a_2+\omega_1)\sin(\phi))\tan(\theta/2) \cr
-((a_1+\omega_2)\cos(\phi)-
(a_2-\omega_1)\sin(\phi))\cot(\theta/2)),
\end{eqnarray}
will be parallel transported, i.e. $U^{'}(\pi) = \frac{\delta^2 \pi}{\delta \tau^2} = 0$. The linear system of 6 real non-homogeneous equation
\begin{eqnarray}\label{46}
\Re(\hat{R}_{00}\eta^0+\hat{R}_{01}\eta^1)&=&\Re( \xi^0 ), \cr
\Im(\hat{R}_{00}\eta^0+\hat{R}_{01}\eta^1)&=&\Im( \xi^0 ), \cr
\Re(\hat{R}_{10}\eta^0+\hat{R}_{11}\eta^1)&=&\Re( \xi^1 ), \cr
\Im(\hat{R}_{10}\eta^0+\hat{R}_{11}\eta^1)&=&\Im( \xi^1 ),\cr
\frac{\delta \theta}{\delta \tau}&=&F_1, \cr
\quad \frac{\delta \phi}{\delta \tau}&=&F_2,
\end{eqnarray}
gives the ``quantum boost" $\vec{a}_Q(\Re(\eta^0),\Im(\eta^0),\Re(\eta^1),\Im(\eta^1),\theta,\phi)$,
and ``quantum rotation"
$\vec{\omega}_Q(\Re(\eta^0),\Im(\eta^0),\Re(\eta^1),\Im(\eta^1),\theta,\phi)$.

The components of the spinor
$(\Re(\eta^0),\Im(\eta^0),\Re(\eta^1),\Im(\eta^1))$ should be agreed with the coefficients $(a,b,c,d)$ into (24) by the condition of solvability of the system (24) reads as $Det|...| = 0$ for the extended matrix $|...|$. Two frequencies $(F_1, F_2)$ will be found from the spectrum of excitations discussed below. One frequency gives the coset deformation acting along some geodesic in $CP(3)$ and the second one gives the velocity of rotation of the geodesic under the action of the gauge isotropy group $H=U(1)\times U(3)$.

\section{Self-interacting quantum electron}
I assume that the spin/charge quantum state of free electron and similar quantum states of freely falling electron should be physically identical. Dynamical equivalence of these quantum states will expressed by the conservation of energy-momentum vector field of quantum electron most naturally may be realized by their affine parallel transport in $CP(3)$. Therefore, if the principle of weak equivalence is valid, then the plane wave cannot be the true quantum state of the free electron and one should derive a new non-linear equation for the description of its ``field-shell". In fact, the requirement of the affine parallel transport puts strong restriction on the ``field-shell" supporting cyclic spin/charge degrees of freedom motion along close geodesic of $CP(3)$.

The local projective coordinates coordinates of eigenstate
\begin{eqnarray}
\pi^i_{(j)}=\begin{cases}
\frac{\psi^i}{\psi^j},& if \quad 1 \leq i < j \cr
\frac{\psi^{i+1}}{\psi^j} & if \quad  j \leq i < 4
\end{cases}
\end{eqnarray}\label{11}
in the map $U_j:\{|\Psi>,|\psi^j| \neq 0 \},1 \leq j \leq 4$
of free electron in $CP(3)$ may be derived from ordinary homogeneous system of eigen-problem. It is easy to see (Leifer, 2007) that under transition from the system of homogeneous equations to the reduced system of non-homogeneous equations (the first equation was omitted)
\begin{eqnarray}
(-E+m c^2) \pi^1+c(p_x+ip_y) \pi^2-c p_z \pi^3 &=& 0 \cr
c(p_x-ip_y) \pi^1-(E+m c^2) \pi^2 &=& cp_z \cr
-cp_z \pi^1-(E+m c^2) \pi^3 &=& c (p_x+ip_y),
\end{eqnarray}\label{13}
one has the single-value solution for eigen-ray
\begin{eqnarray}\label{13}
\pi^1=0, \quad
\pi^2=\frac{-cp_z}{E+mc^2} \quad
\pi^3=\frac{-c(p_x+ip_y)}{E+mc^2},
\end{eqnarray}
in the map $U_1:\{\psi_1 \neq 0\}$ for $E=\sqrt{m^2c^4+c^2p^2}\pm \Delta$.
It is possible only if the determinant of the reduced system
$D=(E^2-m^2c^4-c^2p^2)^2 $ is not equal zero. It is naturally to use these scale-invariant functional variables $(\pi^1,\pi^2,\pi^3)$ in order to establish relation between spin-charge degrees of freedom
and energy-momentum distribution of electron in dynamical space-time (DST) since
the ``off-shell" condition $D=(E^2-m^2c^4-c^2p^2)^2 \neq 0$ opens the way
for self-interaction. New dispersion law will be established
due to formulation of the conservation law of quantum energy-momentum.
In local coordinates (representation) the improper states like plane waves
are simply deleted. It means that trivial free motion of whole quantum
setup in local homogeneous space-time is removed. One may treat this approach as
``self-interacting representation" where only self-interacting effects should be taken into account. Then the dynamics in state space $CP(3)$ plays the leading role whereas a field dynamics in space-time will be derivable.

Infinitesimal energy-momentum  variations  evoked by interaction
charge-spin degrees of freedom (implicit in $\hat{\gamma}^{\mu}$ ) that may be
expressed in terms of local coordinates $\pi^i$ since there is a
diffeomorphism between the space of the rays $CP(3)$ and the $SU(4)$
group sub-manifold of the coset transformations
$G/H=SU(4)/S[U(1) \times U(3)]=CP(3)$ and the isotropy group $H=U(1) \times U(3)$
of some state vector. It will be expressed by the combinations
of the $SU(4)$ generators $\hat{\gamma}_{\mu}$
of unitary transformations that will be defined by an equation arising
under infinitesimal variation of the energy-momentum
\begin{equation}
\Phi_{\mu}^i(\gamma_{\mu}) = \lim_{\epsilon \to 0} \epsilon^{-1}
\biggl\{\frac{[\exp(i\epsilon \hat{\gamma}_{\mu})]_m^i \psi^m}{[\exp(i
\epsilon \hat{\gamma}_{\mu})]_m^j \psi^m }-\frac{\psi^i}{\psi^j} \biggr\}=
\lim_{\epsilon \to 0} \epsilon^{-1} \{ \pi^i(\epsilon
\hat{\gamma}_{\mu}) -\pi^i \},
\end{equation}\label{14}
arose in a nonlinear local realization of $SU(4)$ (Leifer, 2009). Here
$\psi^m, 1\leq m \leq 4$ are the ordinary bi-spinor amplitudes. I calculated the twelve coefficient functions  $\Phi_{\mu}^i(\gamma_{\mu})$ in the map $U_1:\{\psi_1 \neq 0\}$:
\begin{eqnarray}
\Phi_{0}^1(\gamma_{0})&=&0, \quad \Phi_{0}^2(\gamma_{0})=-2i\pi^2,
\quad \Phi_{0}^3(\gamma_{0})=-2i\pi^3; \cr
\Phi_{1}^1(\gamma_{1})&=&\pi^2 -\pi^1 \pi^3,
\quad \Phi_{1}^2(\gamma_{1})=-\pi^1 -\pi^2 \pi^3,
\quad \Phi_{1}^3(\gamma_{1})=-1 -(\pi^3)^2; \cr
\Phi_{2}^1(\gamma_{2})&=&i(\pi^2 +\pi^1 \pi^3),
\quad \Phi_{2}^2(\gamma_{2})=i(\pi^1 +\pi^2 \pi^3),
\quad \Phi_{2}^3(\gamma_{2})=i(-1 +(\pi^3)^2); \cr
\Phi_{3}^1(\gamma_{3})&=&-\pi^3 -\pi^1 \pi^2,
\quad \Phi_{3}^2(\gamma_{3})=-1 -(\pi^2)^2,
\Phi_{3}^3(\gamma_{3})=\pi^1 -\pi^2 \pi^3.
\end{eqnarray}\label{15}

Now I will define the $\Gamma$-vector field
\begin{equation}
\vec{\Gamma}_{\mu}=\Phi_{\mu}^i(\pi^1,\pi^2,\pi^3)\frac{\partial}{\partial \pi^i}
\end{equation}\label{16}
and then the energy-momentum operator will be defined as the \emph{functional
vector field}
\begin{equation}\label{17}
P^{\mu}\vec{\Gamma}_{\mu}\Psi(\pi^1,\pi^2,\pi^3)
= P^{\mu}\Phi_{\mu}^i(\pi^1,\pi^2,\pi^3)
\frac{\partial}{\partial \pi^i}\Psi(\pi^1,\pi^2,\pi^3) + c.c.
\end{equation}
acting on the ``total wave function",
where the ordinary 4-momentum $P^{\mu}=(\frac{E}{c}-\frac{e}{c}\phi,\vec{P} -
\frac{e}{c} \vec{A})=(\frac{\hbar \omega}{c}-\frac{e}{c}\phi,\vec{\hbar k} -
\frac{e}{c} \vec{A})$ (not operator-valued) should be identified with
the solution of quasi-linear ``field-shell" PDE's  for the contravariant
components of the energy-momentum tangent vector field in $CP(3)$
\begin{equation}\label{18}
P^i(x,\pi)=P^{\mu}(x)\Phi_{\mu}^i(\pi^1,\pi^2,\pi^3),
\end{equation}
where $P^{\mu}(x)$ is energy-momentum distribution that comprise of ``field-shell" of the self-interacting electron.

One sees that infinitesimal variation of energy-momentum is represented by
the operator of partial differentiation in complex local coordinates $\pi^i$
with corresponding coefficient functions $\Phi_{\mu}^i(\pi^1,\pi^2,\pi^3)$.
Then the single-component ``total wave function"
$\Psi(\pi^1,\pi^2,\pi^3)$ should be
studied in the framework of new PDE instead of two-component approximation
due to Foldy-Wouthuysen unitary transformations. There are of course four such
functions $\Psi(\pi^1_{(1)},\pi^2_{(1)},\pi^3_{(1)})$, $\Psi(\pi^1_{(2)},
\pi^2_{(2)},\pi^3_{(2)})$,
$\Psi(\pi^1_{(3)},\pi^2_{(3)},
\pi^3_{(3)}), \Psi(\pi^1_{(4)},\pi^2_{(4)},
\pi^3_{(4)})$
- one function in each local map.

The ``field-shell" equations may be derived as the consequence of the
conservation law of the energy-momentum (Leifer, 2009). In the reply on
questions of some colleagues (why, say, Lagrangian is not used for derivation of
the field equations?) I would like to note following. Strictly speaking the least
action principle is realized only in average that is clear
from Feynman's summation of quantum amplitudes. Hence one may suspect that
more deep principle should be used for derivation of fundamental equations of motion. The quantum formulation of the inertia law has been used (Leifer, 2010).

What the inertial principle means for quantum systems and their states?
Formally the inertial principle is tacitly accepted in the package with relativistic invariance. But we already saw that the problem of identification and therefore the localization of quantum particles in classical space-time is problematic and it requires a clarification. I assumed that quantum version of the inertia law may be formulated as follows:

\textbf{inertial quantum motion of quantum system may be expressed as a
 self-conservation of its local dynamical variables like energy-momentum, spin, charge, etc.}

The conservation law of the energy-momentum vector field in $CP(3)$ during
inertial evolution will be expressed by the equation of the affine parallel transport
\begin{equation}
\frac{\delta [P^{\mu}\Phi_{\mu}^i(\gamma_{\mu})]}{\delta \tau}=0,
\end{equation}\label{32}
which is equivalent to the following system of four coupled quasi-linear PDE
for dynamical space-time distribution of energy-momentum ``field-shell" of
quantum state
\begin{equation}
V^{\mu}_Q (\frac{\partial
P^{\nu}}{\partial x^{\mu} } + \Gamma^{\nu}_{\mu \lambda}P^{\lambda})=
-\frac{c}{\hbar}(\Gamma^m_{mn} \Phi_{\mu}^n(\gamma) + \frac{\partial
\Phi_{\mu}^n (\gamma)}{\partial \pi^n}) P^{\nu}P^{\mu},
\end{equation}\label{33}
and ordinary differential equations for relative amplitudes
\begin{equation}
\quad \frac{d\pi^k}{d\tau}= \frac{c}{\hbar}\Phi_{\mu}^k P^{\mu},
\end{equation}\label{34}
which is in fact the \emph{equations of characteristic} for linear ``super-Dirac"
equation
\begin{equation}\label{35}
i P^{\mu}\Phi_{\mu}^i(\gamma_{\mu})\frac{\partial \Psi}{\partial \pi^i} =mc \Psi
\end{equation}
that supposes ODE for single ``total state function"
\begin{equation}\label{36}
i \hbar \frac{d \Psi}{d \tau} =mc^2 \Psi
\end{equation}
with the solution for variable mass $m(\tau)$
\begin{equation}\label{37}
\Psi(T) = \Psi(0) e^{-i\frac{c^2 }{\hbar}\int_0^T m(\tau) d\tau}.
\end{equation}

Probably, simple
relation given by Fubini-Study metric for the square of the frequency
\begin{eqnarray}
\Omega^2=(\frac{dS_{F.-S.}}{d\tau})^2=G_{ik*}\frac{d\pi^i}{d\tau}\frac{d\pi^{k*}}{d\tau}
=\frac{c^2}{\hbar^2}G_{ik*}(\Phi^i_{\mu}P^{\mu})(\Phi^{k*}_{\nu}P^{\nu})
\end{eqnarray}\label{115}
associated with velocity traversing geodesic line (6.14)
during spin/charge variations in $CP(3)$ sheds the light on the mass problem
of self-interacting electron in the connection with action principle on $CP(3)$.
Taking into account the ``off-shell" dispersion law
\begin{eqnarray}
\frac{\hbar^2}{c^2} G_{ik*}\frac{d\pi^i}{d\tau}\frac{d\pi^{k*}}{d\tau}
=G_{ik*}(\Phi^i_{\mu}P^{\mu})(\Phi^{k*}_{\nu}P^{\nu*}) \cr
 =(G_{ik*}\Phi^i_{\mu}\Phi^{k*}_{\nu}) P^{\mu} P^{\nu*}
= g_{\mu \nu}P^{\mu}P^{\nu*}=m^2 c^2
\end{eqnarray}\label{115}
one has
\begin{eqnarray}
i \frac{d \Psi}{d \tau} = \frac{mc^2}{\hbar} \Psi = \Psi \sqrt{G_{ik*}\frac{d\pi^i}{d\tau}\frac{d\pi^{k*}}{d\tau}} =\Psi \sqrt{dS^2_{F.-S.}}=\pm \Psi dS_{F.-S.}
\end{eqnarray}\label{116}
and, therefore,
\begin{equation}\label{37}
\Psi(T) = \Psi(0) e^{\pm i S_{F.-S.}}.
\end{equation}
In other words
\begin{equation}\label{37}
dS_{F.-S.}= \frac{c^2}{\hbar}m d\tau.
\end{equation}
The metric tensor of the local DST in the vicinity of electron
$g_{\mu \nu} = G_{ik*}\Phi^i_{\mu}\Phi^{k*}_{\nu}$ is state dependable and it will be discussed in a separated article.

The system of quasi-liner PDE's (6.10) following from the conservation law, ODE's
and algebraic linear non-homogeneous equations
comprise of the self-consistent problem for stability (in fact - existing) of electron. Their solution have been shortly discussed
(Leifer, 2009). The theory of these equations is well known (Courant \& Hilbert, 1989). One has the quasi-linear PDE system with identical principle part $V^{\mu}_Q$
for which we will build system of ODE's of characteristics
\begin{eqnarray}
\frac{d x^{\nu}}{d \tau}&=&V^{\nu}_Q,\cr
\frac{d P^{\nu}}{d \tau}&=&-V^{\mu}_Q
\Gamma^{\nu}_{\mu \lambda}P^{\lambda}-\frac{c}{\hbar}(\Gamma^m_{mn}
\Phi_{\mu}^n(\gamma) + \frac{\partial
\Phi_{\mu}^n (\gamma)}{\partial \pi^n}) P^{\nu}P^{\mu}, \cr
\frac{d\pi^k}{d\tau} &=& \frac{c}{\hbar}\Phi_{\mu}^k P^{\mu}.
\end{eqnarray}\label{44}
One of the integrable combination is as follows
\begin{eqnarray}
\frac{\delta x^0}{V^0_Q}=\frac{\delta P^0}{P^0(L_0P^0+L_1P^1+L_2P^2+L_3P^3)},
\end{eqnarray}
where $L_{\mu}=-\frac{c}{\hbar}(\Gamma^m_{mn} \Phi_{\mu}^n(\gamma) + \frac{\partial
\Phi_{\mu}^n (\gamma)}{\partial \pi^n})$. If $L_0P^0<0$ then one has implicit solution
\begin{eqnarray}
\frac{x^0}{a_{\alpha}x^{\alpha}} + t^0=-\frac{2}{L_{\alpha}P^{\alpha}}\tanh^{-1}(1+\frac{2L_0P^0}{L_{\alpha}P^{\alpha}}),
\end{eqnarray}
where $t^0$ is an integration constant. An explicit solution for energy is the kink
\begin{eqnarray}
P^0=\frac{L_{\alpha}P^{\alpha}}{2L_0}[ \tanh(-(\frac{x^0}{a_{\alpha}x^{\alpha}} + t^0)\frac{L_{\alpha}P^{\alpha}}{2})-1].
\end{eqnarray}
If I put $\frac{L_{\alpha}P^{\alpha}}{2}=1, L_0=1, V=a_{\alpha}x^{\alpha}=0.6$  the kink solution may be represented by the graphic in Fig. 2.
\begin{figure}[h]
  \begin{center}
    \includegraphics[width=4in]{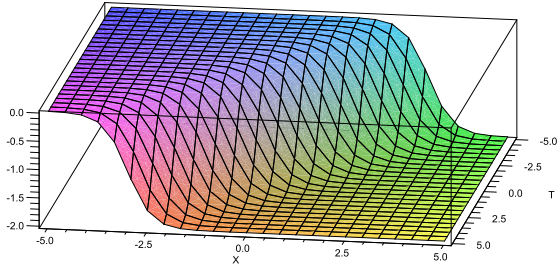}\\
  \caption{The kink solution of the quasi-linear PDE's in dynamical space-time showing the distribution of energy-momentum ``field-shell" of extended quantum electron. It is not solution of a runaway type.}
  \end{center}
  \end{figure}
This solution represent the lump of self-interacting electron through an electro-magnetic-like field in the co-moving Lorentz reference frame. The nature of this field will be discussed in a separate article. In the standard QED self-interacting effects are treated as a polarization of the vacuum. In the present picture the lump is dynamically self-supporting by outward and inward waves whose characteristics are represented by the equations (6.21).

\section{Stability of energy-momentum characteristics and dispersion law}
Let me discuss the system of the characteristics
\begin{eqnarray}
\frac{dP^{\nu}}{d \tau}&=&-V^{\mu}_Q  \Gamma^{\nu}_{\mu \lambda}P^{\lambda}-\frac{c}{\hbar}(\Gamma^m_{mn} \Phi_{\mu}^n(\gamma) + \frac{\partial
\Phi_{\mu}^n (\gamma)}{\partial \pi^n}) P^{\nu}P^{\mu}.
\end{eqnarray}
Their stability will provide the equilibrium of the distribution of energy-momentum in DST. The standard approach to stability analysis instructs us to find the stationary points. The stationary condition
\begin{eqnarray}
\frac{\delta P^{\lambda}}{\delta \tau}&=&0
\end{eqnarray}
leads to the system of algebraic equations
\begin{eqnarray}
V^{\mu}_Q  \Gamma^{\nu}_{\mu \lambda}P^{\lambda}+\frac{c}{\hbar}(\Gamma^m_{mn} \Phi_{\mu}^n(\gamma) + \frac{\partial
\Phi_{\mu}^n (\gamma)}{\partial \pi^n}) P^{\nu}P^{\mu}=0.
\end{eqnarray}
The probing solution in the vicinity of the stationary points $P^{\mu}_0$ is as follows
\begin{eqnarray}
P^{\mu}(\tau)=P^{\mu}_0 + p^{\mu} e^{\omega \tau}.
\end{eqnarray}
This solution being substituted in the equation (7.3)
gives the homogeneous linear system
\begin{eqnarray}
\frac{\hbar \omega}{c}p^{\nu}+\frac{\hbar }{c}V^{\mu}_Q  \Gamma^{\nu}_{\mu \lambda}p^{\lambda}+ (\Gamma^m_{mn} \Phi_{\mu}^n(\gamma) + \frac{\partial
\Phi_{\mu}^n (\gamma)}{\partial \pi^n}) p^{\mu}P^{\nu}_0 = 0.
\end{eqnarray}
The determinant of this system is as follows
\begin{eqnarray}
D_1= (\frac{\hbar \omega}{c})^4+\alpha (\frac{\hbar \omega}{c})^3+\beta (\frac{\hbar \omega}{c})^2+\gamma (\frac{\hbar \omega}{c})+\delta,
\end{eqnarray}
with complicated coefficients $\alpha, \beta, \gamma, \delta$. I put $K^{\nu}_{\lambda}=\frac{\hbar }{c}V^{\mu}_Q  \Gamma^{\nu}_{\mu \lambda}$ and $M^{\nu}_{\mu}=(\Gamma^m_{mn} \Phi_{\mu}^n(\gamma) + \frac{\partial
\Phi_{\mu}^n (\gamma)}{\partial \pi^n}) P^{\nu}_0$ then
one may find that
\begin{eqnarray}
\alpha=Tr(K^{\nu}_{\lambda})+ Tr(M^{\nu}_{\mu})
\end{eqnarray}
and
\begin{eqnarray}
\beta=K^0_0(L_1P_0^1+L_2P_0^2+L_3P_0^3)+K^1_1(L_0P_0^0+L_2P_0^2+L_3P_0^3) \cr +K^2_2(L_1P_0^1+L_0P_0^0+L_3P_0^3)+K^3_3 (L_1P_0^1+L_0P_0^0+L_2P_0^2)\cr
-K^0_1 L_0P_0^1-K^1_0 L_1P_0^0 -K^0_2 L_0P_0^2-K^2_0 L_2P_0^0
-K^0_3 L_0P_0^3-K^3_0 L_3P_0^0 \cr -K^1_2 L_1P_0^2-K^2_1 L_2P_0^1
-K^1_3 L_1P_0^3-K^3_1 L_3P_0^1 -K^2_3 L_2P_0^3-K^3_2 L_3P_0^2,
\end{eqnarray}
whereas $ \gamma, \delta$ have higher order in $\frac{\hbar }{c}$ and they may be temporarily  discarded in approximate dispersion law.
This dispersion law may be written as follows
\begin{eqnarray}
 (\frac{\hbar \omega}{c})^2[(\frac{\hbar \omega}{c})^2+\alpha (\frac{\hbar \omega}{c})+\beta] =0.
\end{eqnarray}
The trivial solution $\omega_{1,2}=0$ has already been discussed (Leifer, 2009). Two non-trivial solutions when $\alpha^2 \gg \beta$ are given by the equations
\begin{eqnarray}
 \hbar \omega_{3,4} =c \alpha \frac{-1 \pm \sqrt{1-\frac{4\beta}{\alpha^2}}}{2} \approx c \alpha \frac{-1 \pm (1-\frac{2\beta}{\alpha^2})}{2};\cr
 \hbar \omega_3=\frac{-c \beta}{\alpha}, \quad \hbar \omega_4=-c \alpha+\frac{c \beta}{\alpha}.
 \end{eqnarray}
Both parameters $\alpha, \beta$ are in fact complex functions of $(\pi^1,\pi^2,\pi^3)$ but they are linear in stationary momenta $P^{\mu}_0$.
Coefficients $K^{\nu}_{\mu} \propto \frac{\hbar }{c}$ are much smaller than $M^{\nu}_{\mu}$. The  negative real part of these two roots substituted in the probing function (7.4) will define attractors and two finite masses.

One should find solution of the non-linear system (7.3).
Its approximate solution in the vicinity of $P^{\mu}_{test} = (mc^2, 0, 0, 0) $
has been found by the method of Newton:
\begin{eqnarray}
P_0^{\mu} =P^{\mu}_{test} + \delta^{\mu}+...,
\end{eqnarray}\label{55}
where $\delta^{\mu}$ is the solution of the Newton's first approximation equations
\begin{eqnarray}
(2L_0mc+K^0_0)\delta^0+(L_1mc+K^0_1)\delta^1+\cr
(L_2mc+K^0_2)\delta^2+(L_3mc+K^0_3)\delta^3 &=& -\frac{(L_0m^2c^4+K^0_0mc^3)}{c^2} \cr
K^1_0\delta^0+(L_0mc+K^1_1)\delta^1+K^1_2\delta^2+K^1_3\delta^3 &=& -K^1_0mc \cr
K^2_0\delta^0+K^2_1\delta^1+(L_0mc+K^2_2)\delta^2+K^2_3\delta^3 &=& -K^2_0mc \cr
K^3_0\delta^0+K^3_1\delta^1+K^3_2\delta^2+(L_0mc+K^3_3)\delta^3 &=& -K^3_0mc,
\end{eqnarray}\label{56}
where $L_{\mu}=(\Gamma^m_{mn} \Phi_{\mu}^n(\gamma) + \frac{\partial
\Phi_{\mu}^n (\gamma)}{\partial \pi^n})$ is now dimensionless.
The solution of this system is as follows:
\begin{eqnarray}
\delta^0 &=& -mc^2 \frac{mcL_0^2-K^1_0L_1-K^2_0L_2- K^3_0L_3}{2mcL_0^2-K^1_0L_1-K^2_0L_2- K^3_0L_3}\cr
\delta^1 &=& -mc \frac{K^1_0L_1}{2mcL_0^2-K^1_0L_1-K^2_0L_2- K^3_0L_3}\cr
\delta^2 &=& -mc \frac{K^2_0L_2}{2mcL_0^2-K^1_0L_1-K^2_0L_2- K^3_0L_3}\cr
\delta^3 &=& -mc \frac{K^3_0L_3}{2mcL_0^2-K^1_0L_1-K^2_0L_2- K^3_0L_3},
\end{eqnarray}\label{78}
where I took into account $K_{\nu}^{\mu}+K_{\mu}^{\nu}=0$.

If hypothesis about dynamical nature of electron mass defined by self-interacting
spin/charge degrees of freedom is correct then it is very natural to assume that
\begin{eqnarray}\label{61}
F_1 = \frac{\delta \theta}{\delta \tau} =\Re(\omega_{3}) =
\frac{c}{\hbar}\Re(\frac{- \beta}{ \alpha}),or \cr
F_1 = \frac{\delta \theta}{\delta \tau} =\Re(\omega_{4}) =
\frac{c}{\hbar} \Re(- \alpha+\frac{ \beta}{\alpha}), and \cr
F_2 = \frac{\delta \phi}{\delta \tau} =\Im(\omega_{3})
=\frac{c}{\hbar}\Im(\frac{- \beta}{ \alpha}),or \cr
F_2 = \frac{\delta \phi}{\delta \tau} =\Im(\omega_{4})
=\frac{c}{\hbar} \Im(- \alpha+\frac{ \beta}{\alpha}).
\end{eqnarray}
The conditions of instability are given by the simple inequalities:
$\Re(\alpha) \Re(\beta) + \Im(\alpha) \Im(\beta) < 0 $ for $\omega_3$ and
$\Re(\alpha) \Re(\beta) + \Im(\alpha) \Im(\beta) - \Re(\alpha) |\alpha|^2 > 0$ for
$\omega_4$.
However, the solution of complicated self-consistent problem (5.8), (6.21), (7.9) and (7.13), (7.14) is not found up to now. Therefore the problem of dynamically generated mass is not solved yet. But the ``field-shell" of the self-interacting quantum electron may be interesting in the connection with old problem of the runaway solution (see (Hammond, 2010) and references therein). Contradictable field structure of quantum electron requires some clarifications.

Let me start with simplified equation (7.1) in the vicinity of the point $(\pi^1 = \pi^2 = \pi^3 = 0 )$. The equation of characteristics will be treated here as the equations of motion for the self-momentum
\begin{eqnarray}
\frac{dP^{\nu}}{d \tau}=-\frac{c}{\hbar}K^{\nu}_{\mu}P^{\mu} +\frac{4ic}{\hbar} P^0P^{\nu},
\end{eqnarray}
where the linear term contains the matrix
\begin{eqnarray}
K^{\nu}_{\mu}=\frac{\hbar}{c^2}\vec{a}\vec{x} \left(
\begin{matrix}
0 & -a_1 & -a_2 & -a_3 \cr
a_1 & 0 & -\omega_3 & \omega_2 \cr
a_2 & \omega_3 & 0 & -\omega_1  \cr
a_3 & -\omega_2 & \omega_1 & 0
\end{matrix}
\right)
\end{eqnarray}
and the non-linear term is imaginary since $L_{\mu}(0)=(-4i,0,0,0)$.
The right part linearized in $P^{\mu}$ is the Jacobi matrix
\begin{eqnarray}
J^{\nu}_{\mu}=\left(
\begin{matrix}
\frac{8ic}{\hbar} P^0 & -K^0_1 & -K^0_2 & -K^0_3 \cr
-K^1_0+\frac{4ic}{\hbar} P^1 & \frac{4ic}{\hbar} P^0 & -K^1_2 & -K^1_3 \cr
-K^2_0+\frac{4ic}{\hbar} P^2 & -K^2_1 & \frac{4ic}{\hbar} P^0 & -K^2_3  \cr
-K^3_0+\frac{4ic}{\hbar} P^3 & -K^3_1  & -K^3_2  & \frac{4ic}{\hbar} P^0
\end{matrix}
\right)
\end{eqnarray}
computed at the stationary point (7.11). If one puts $P^{\mu}_0 = (0,0,0,0)$,
and for simplicity  $a_2=a_3=\omega_2=\omega_3 = 0$ then one has a neutral stability since eigen values of $J^{\nu}_{\mu}$ are $\lambda_1=i\omega_1, \lambda_2=-i\omega_1,\lambda_3=ia_1, \lambda_4=-ia_1$.
If $P^{\mu}_0 = P^{\mu}_{test} = (mc^2,0,0,0)$ then eigen values will be as follows:
$\lambda_1=i(6m+\sqrt{4m^2+a_1^2}), \lambda_2=i(6m-\sqrt{4m^2+a_1^2}),\lambda_3=i(4m+\omega_1), \lambda_4=i(4m-\omega_1)$.
If, however, one takes into account the first approximation corrections (7.13) then one get  $\lambda_1=i[6(m+\delta^0)+\sqrt{4(m+\delta^0)^2+a_1^2-4ia_1\delta^1}], \lambda_2=i[6(m+\delta^0)-\sqrt{4(m+\delta^0)^2+a_1^2-4ia_1\delta^1}],
\lambda_3=i(4m+\omega_1+\delta^0), \lambda_4=i(4m-\omega_1+\delta^0)$
showing possible bifurcation instability if the mass $m$ used as a bifurcation parameter.

Analysis of the full Riccati system (7.1) and even its linearized version is too complicated and it is not finished. One may, however, find the exact solutions of (7.15) in two separated cases: $(a_1 \neq 0, a_2=a_3=\omega_1 =\omega_2=\omega_3 = 0)$ and $(\omega_1 \neq 0, a_1=a_2=a_3=\omega_2=\omega_3 = 0)$ where I assumed that $(a_1 \neq 0,\omega_1 \neq 0)$ are not the field solution of the full self-consistent problem but simply constant number parameters (I've put $\vec{a}\vec{x}=1$). Even oversimplified results given merely for illustration show a rich internal dynamical structure of self-interacting electron.

Exact solution of (7.15) will be represented graphically with needed explanations.

A. Let me assume $(a_1 = 2, a_2=a_3=\omega_1 =\omega_2=\omega_3 = 0)$, i.e. $x=1/2$. The boost parameter $a_1$ has the dimension of a momentum here, and four integration constants have following physical dimensions: $[C1=3]=momentum, [C2=1]=momentum, [C3=5]=dimensionless, [C4=3]=dimensionless $.
\begin{eqnarray}
P^0(\tau)=\frac{ia_1[C_4\cos(\frac{a_1 c \tau}{\hbar})-\sin(\frac{a_1 c \tau}{\hbar})]}{4[C_3+C_4\sin(\frac{a_1 c \tau}{\hbar})-\cos(\frac{a_1 c \tau}{\hbar})]}
\end{eqnarray}
\begin{figure}[h]
  \begin{center}
    \includegraphics[width=2in]{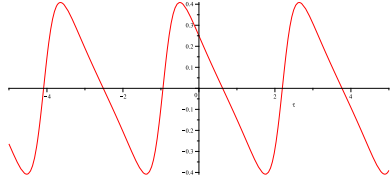}\\
  \caption{The imaginary part of the energy $P^0(\tau)$ under the boost $a_1=2$ in x-direction.}
  \end{center}
  \end{figure}
\begin{eqnarray}
P^3(\tau)=\frac{C_2}{C_3+C_4\sin(\frac{a_1 c \tau}{\hbar})-\cos(\frac{a_1 c \tau}{\hbar})}
\end{eqnarray}
  \begin{figure}[h]
  \begin{center}
    \includegraphics[width=2in]{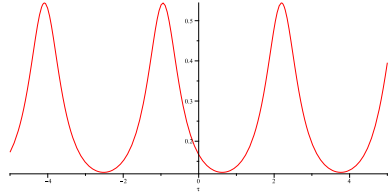}\\
  \caption{The real part of the z-momentum $P^3(\tau)$ under the boost $a_1=2$ in x-direction.}
  \end{center}
  \end{figure}
One sees that the boost of the Lorentz frame attached to GCS of the electron leads to periodic oscillations of all components of energy-momentum.

B. Let me assume now $(\omega_1 = 1, a_1=a_2=a_3=\omega_2=\omega_3 = 0)$.
The rotation parameter $\omega_1 $ has here the dimension of a momentum, and four integration constants have following physical dimensions: $[C1=3]=momentum, [C2=1]=momentum^{-1}, [C3=5]=action, [C4=3]=action$.
\begin{eqnarray}
P^0(\tau)_{re}=\Re(\frac{\hbar}{C_2 \hbar - 4ic\tau })
\end{eqnarray}
\begin{figure}[h]
  \begin{center}
    \includegraphics[width=2in]{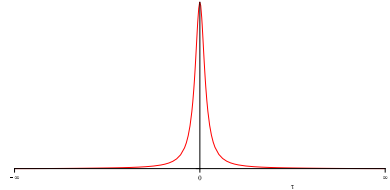}\\
  \caption{The real part of the energy $P^0(\tau)_{re}$ under the rotation $\omega_1 = 1$ about axes $OX$.}
  \end{center}
  \end{figure}
\begin{eqnarray}
P^0(\tau)_{im}=\Im(\frac{\hbar}{C_2 \hbar - 4ic\tau})
\end{eqnarray}
\begin{figure}[h]
  \begin{center}
    \includegraphics[width=2in]{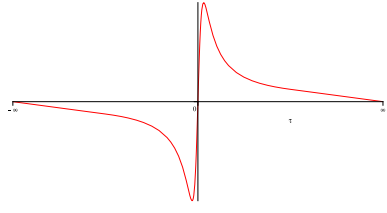}\\
  \caption{The imaginary part of the energy $P^0(\tau)_{im}$ under the rotation $\omega_1 = 1$ about axes $OX$.}
  \end{center}
  \end{figure}
\begin{eqnarray}
P^2(\tau)_{re}=\Re(-\frac{C_3\sin(\frac{\omega_1 c \tau}{\hbar})+ C_4\cos(\frac{\omega_1 c \tau}{\hbar})}{C_2 \hbar - 4ic\tau })
\end{eqnarray}
  \begin{figure}[h]
  \begin{center}
    \includegraphics[width=2in]{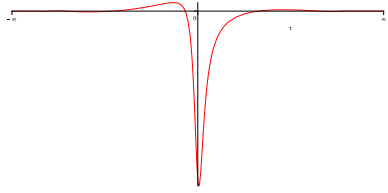}\\
  \caption{The real part of the y-momentum $P^2(\tau)_{re}$ under the rotation $\omega_1 = 1$ about axes $OX$.}
  \end{center}
  \end{figure}
  \begin{eqnarray}
P^2(\tau)_{im}=\Im(-\frac{C_3\sin(\frac{\omega_1 c \tau}{\hbar})+ C_4\cos(\frac{\omega_1 c \tau}{\hbar})}{C_2 \hbar - 4ic\tau })
\end{eqnarray}
  \begin{figure}[h]
  \begin{center}
    \includegraphics[width=2in]{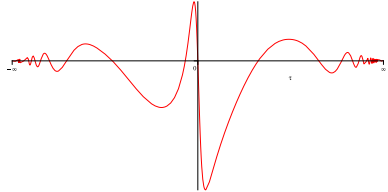}\\
  \caption{The imaginary part of the y-momentum $P^2(\tau)_{im}$ under the rotation $\omega_1 = 1$ about axes $OX$.}
  \end{center}
  \end{figure}
One sees that rotations of the Lorentz frame attached to GCS of the electron describe the radiation force leading to irreversible loss of energy and momentum. Generally, the oversimplified equation (7.15) leads to finite solutions for all components of the energy-momentum, i.e. no runaway solutions for self-interacting electron. I suspect that self-consistent solution may lead only to more fast decay of the momentum as a function of deviation from stationary state. In the previous examples the ``proper quantum time" $\tau$ is in fact the measure of such deviation in $CP(3)$.

\section{Conclusion}
Analysis of the localization problem insists to make attempts of the \emph{intrinsic unification} of quantum principles based on the fundamental concept of quantum amplitudes and the principle of relativity ensures the physical equivalence of any conceivable quantum setup. The realization of such program evokes the necessity of the state-dependent affine gauge field in the state space that acquires reliable physical basis under the quantum formulation of the inertia law (self-conservation of local dynamical variables of quantum particle during inertial motion). Representation of such affine gauge field in dynamical space-time has been applied to the relativistic extended self-interacting Dirac's electron (Leifer, 2009,2010). Thus one has unconstrainedly given localized (soliton-like) solution and promising dispersion law with a mass gap.

The simplest formulation of the quantum inertia law by the affine parallel transport of energy-momentum in the projective Hilbert state space has been proposed  (Leifer, 2010). Such approach paves the way to clarification of the old problem of inertial mass and such ``fictitious" forces as, say, centrifugal force. Shortly speaking the inertia and inertial forces are originated not in space-time but it the space of quantum states since they are generated by the deformation of quantum states as a reaction on an external influence.

 \label{lastpage}
\end{document}